\documentclass[jkps,preprint,fleqn,showkeys]{revtex4}
\usepackage[pdftex]{graphicx}
\usepackage{amssymb}
\usepackage{amsmath}
\usepackage{graphicx}
\usepackage{multirow}
\usepackage[autostyle]{csquotes}

\usepackage{bm}

\newcommand{\nn}{\nonumber}

\newcommand{\rr}{{\gamma\gamma}}

\newcommand{\pb}{{\;{\rm pb}}}
\newcommand{\fb}{{\;{\rm fb}}}

\newcommand{\gev}{{\;{\rm GeV}}}

\newcommand{\beq}{\begin{equation}}
\newcommand{\eeq}{\end{equation}}
\newcommand{\bea}{\begin{eqnarray}}
\newcommand{\eea}{\end{eqnarray}}
\newcommand{\barr}{\begin{array}}
\newcommand{\earr}{\end{array}}
\newcommand{\bc}{\begin{center}}
\newcommand{\ec}{\end{center}}
\newcommand{\bit}{\begin{itemize}}
\newcommand{\eit}{\end{itemize}}
\newcommand{\ben}{\begin{enumerate}}
\newcommand{\een}{\end{enumerate}}

\newcommand{\dt}{\delta}

\newcommand{\sg}{\sigma}

\newcommand{\gm}{\gamma}

\newcommand{\wpm}{W^\pm}

\newcommand{\pt}{p_T}

\newcommand{\ttop}      {{t\bar{t}}}
\newcommand{\ttt}      {\overset{\text{\fontsize{2pt}{2pt}\selectfont(--)}}{t}}
\newcommand{\bb}      {{b \bar{b}}}
\newcommand{\ww}      {{W^+ W^-}}

\newcommand{\eq}[1]{Eq.~(\ref{#1})}

\begin{document}
\setcounter{page}{0}
\title[]{A Panoramic Study of $K$-Factors for 111 Processes \\
at the 14 TeV LHC}
\author{Dongjoo \surname{Kim}}
\author{Soojin  \surname{Lee}}
\author{Hanseok  \surname{Jung}}
\author{Dongchan  \surname{Kim}}
\author{Jinheung  \surname{Kim}}
\author{Jeonghyeon  \surname{Song}}
\email{jhsong@konkuk.ac.kr}
\affiliation{Department of Physics, Konkuk University, Seoul 05029, Republic of Korea}

\date[]{Received March 2024}

\begin{abstract}
In this comprehensive study, we investigate $K$-factors ($K=\sigma_{\text{NLO}}/\sigma_{\text{LO}}\equiv 1+\delta K$) for a broad array of Standard Model processes at the 14 TeV LHC, which are pivotal for background assessments in Beyond the Standard Model (BSM) searches. Using {\sc\small MadGraph5\_aMC@NLO}, we calculate the leading-order and next-to-leading order (NLO) cross-sections and compute the corresponding $K$-factors for 111 processes. Our analysis reveals $K$-factors ranging from 1.005 for $pp \to jjj$ to 4.221 for $pp\to W^\pm \gamma\gamma\gamma$. Key findings include: (i) processes involving photons display significantly high $K$-factors, attributed to gluon-initiated processes at NLO; (ii) processes with multiple particle productions, particularly those involving vector bosons, exhibit elevated $K$-factors due to multiple real emission processes; (iii) there exists an inverse correlation between the number of jets and $\delta K$, indicating that the addition of jets generally leads to a decrease in $\delta K$. Additionally, our investigation into differential $K$-factors relative to transverse momentum and invariant mass shows notable increases with higher $p_T$, but minimal changes with invariant mass. This study highlights the indispensable role of precise $K$-factor evaluations for accurate interpretations of BSM search outcomes.
\end{abstract}


\maketitle

\section{Introduction}

The Standard Model (SM) of particle physics has been robustly validated through numerous experiments, 
culminating in the landmark discovery of a Higgs boson with a mass of $125\gev$ at the LHC\cite{ATLAS:2012yve,CMS:2012qbp}. 
Despite these achievements, the search for Beyond the Standard Model (BSM) theories persists, 
driven by unresolved questions such as the nature of dark matter\cite{Navarro:1995iw,Bertone:2004pz}, the origins of neutrino masses, the metastability of the SM vacuum\cite{Degrassi:2012ry}, and the naturalness problem\cite{Dimopoulos:1995mi,Chan:1997bi,Craig:2015pha}. 
The lack of new signals akin to the discovery of the $J/\psi$ particle\cite{SLAC-SP-017:1974ind,E598:1974sol}
suggests that BSM signals may be rare or hidden within complex particle processes, 
underscoring the importance of a comprehensive assessment of potential SM backgrounds 
to unearth promising BSM discovery channels.

Evaluating SM backgrounds typically relies on leading order (LO) cross sections. 
Yet, in certain cases, the next-to-leading order (NLO) corrections are notably high, indicating significant departures from the expected perturbative behavior.  
These discrepancies can have a profound impact on BSM searches\cite{Hankele:2007sb,Campanario:2013mha}, especially in scenarios where both signals and backgrounds are rare\cite{Jueid:2023fgo}.

The $K$-factor, representing the ratio of NLO or higher-order to LO cross sections, plays a pivotal role in enhancing the precision of background estimations in BSM searches. This methodology has proven to be particularly influential in studies of single Higgs production, where incorporating higher-order corrections has been crucial for comparing theoretical predictions  with experimental data\cite{Spira:1997dg,Anastasiou:2002yz,Ahrens:2009cxz,Anastasiou:2015vya,Spira:2016ztx}.

Calculating observables beyond the Born approximation in Quantum Chromodynamics (QCD)  presents considerable challenges, such as the complexity of computing virtual corrections, the issue of infrared divergences, and the integration of these elements with parton showers.
The development of {\sc\small MadGraph\_aMC@NLO} (MG@NLO) has  facilitated the automated computation of NLO cross sections, addressing these challenges with significant efficiency\cite{Alwall:2014hca}. 
However, the extensive computational resources needed to generate large NLO datasets continue to be an obstacle for rapid SM background estimation. 
Consequently, providing $K$-factors across a broad range of SM processes emerges as an invaluable resource.

Previous studies have thoroughly documented the LO and NLO cross sections for a variety of SM processes at the 13 TeV LHC\cite{Alwall:2014hca,Ghosh:2022lrf}. 
Building on this foundation, our research extends these analyses to the 14 TeV collision energy, aligning with the forthcoming High-Luminosity LHC phase. 
We focus on identifying patterns in processes with notably high $K$-factors, 
especially those involving photons and multiple particle productions. 
A key objective is to clarify the underlying reasons for these elevated $K$-factors.

Additionally, we explore how $K$-factors vary with kinematic variables like transverse momentum ($p_T$) and invariant mass. 
Given that the BSM search usually targets specific parameter spaces, 
differential $K$-factors play a pivotal role in distinguishing signals from backgrounds. 
Notably, regions of high $p_T$, which can be precisely investigated thanks to the LHC's high luminosity, hold promise for the detection of BSM signals.
Moreover, understanding the variation of $K$-factors with invariant mass is essential for the accurate reconstruction of new particle masses. 
While previous studies have concentrated on particular processes for differential $K$-factors\cite{Alioli:2010xd,Frixione:2014qaa,Zhang:2014gcy,Kallweit:2015dum,Frixione:2015zaa,
Czakon:2017wor},  our comparative study across various processes seeks to reveal the universal behaviors of $K$-factors concerning $p_T$ and invariant mass.
 
 This paper is structured as follows: 
 We begin with Section~\ref{sec:global:K}, investrigating global $K$-factors for 111 SM processes, 
 derived from both LO and NLO cross-section calculations. 
 Section~\ref{sec:panoramic} shifts the focus to processes with $K$-factors above 1.5, 
 particularly highlighting those involving photons. 
 The study of differential $K$-factors in relation to transverse momentum and invariant mass is the focus of Section~\ref{sec:pt:invmass}. Finally, we conclude in Section~\ref{sec:conclusions}.
 
\section{$K$-factors for total rates}
\label{sec:global:K}

The $K$-factor is the ratio of the cross-section calculated at a higher order in perturbation theory 
to that calculated at LO,
quantifying the impact of higher-order corrections on the considered process. 
 In this section, we calculate $K$-factors for 111 processes at the 14 TeV LHC, expected to serve as backgrounds for most BSM searches. 
For the comparative study of these 111 processes, we define the $K$-factor as the ratio of the NLO (in QCD) cross-section to LO cross-sections:
\begin{equation}
\label{eq:K:factor}
K \equiv 1 + \delta K = \frac{\sigma_{\text{NLO}}}{\sigma_{\text{LO}}}.
\end{equation}
For the NLO cross section calculation, we used Fixed Order method in MG@NLO,
termed as fNLO in Ref.\cite{Alwall:2014hca}.

The primary physics parameters and definitions for final-state objects used in our computation 
are outlined as follows:
\begin{itemize}
\item The Higgs boson and top quark masses are set as:
\begin{equation}
m_H = 125\gev, \quad m_t = 173 \gev.
\end{equation}
\item For the parton distribution function set, 
we use {\sc\small NNPDF31\_lo\_as\_0118}\cite{NNPDF:2017mvq} in the four-flavor scheme.
\item For the renormalization scale $\mu_R$ and the factorization scale $\mu_F$, we adopt:
\begin{equation}
\label{eq:scale:def}
\mu_R = \mu_F = \frac{H_T}{2} \equiv \mu_0,
\end{equation}
where $H_T$ is the scalar sum of the transverse masses of all final-state particles, defined by:
\begin{equation}
H_T = \sum_i \sqrt{(\pt^i)^2 + m_i^2}.
\end{equation}
\item A diagonal CKM matrix is assumed.
\item Jets are clustered using the $k_t$ algorithm\cite{Catani:1993hr,Ellis:1993tq} with $R=0.7$, $\pt^j > 30 \gev$, and $|\eta_j| < 4$.
\item Photons are reconstructed through the Frixione isolation\cite{Frixione:1998jh} 
with $R=0.4$, $\pt^\gamma > 20 \gev$, and $|\eta_\gamma| < 2.5$.
\end{itemize}

We acknowledge significant uncertainties stemming from the renormalization and factorization scales in calculating total cross-sections, noting that uncertainties from PDFs and integration errors are comparatively minor, typically on the order of a few percent. The scale uncertainty in the LO cross-section varies significantly across different processes. 
The production of $\wpm$ or $Z$, denoted to be $V = W^\pm, Z$, illustrates this feature, with scale uncertainty around 15\% for single $V$ production and decreasing to below 1\% for three $V$ production. 

Photon production processes exhibit higher uncertainties. For example, the uncertainty in the LO cross-section for $pp \to \gamma W^\pm$ is roughly double that for $pp \to ZW^\pm$. Including jets further amplifies scale uncertainties. Adding a single jet to single $V$ production increases the uncertainty by about 20\%, compared to the case without a jet. The uncertainty escalates about 40\% with the inclusion of three jets. This trend peaks in jet productions without an electroweak gauge boson:
specifically, the scale uncertainty can reach up to about 60\% for $pp\to b\bar{b}jj$.

Our focus, however, is on NLO cross sections, which generally show reduced scale dependence 
due to the inclusion of higher-order corrections in quantum field theory calculations. 
This reduction is particularly significant in processes with large scale uncertainty at LO. 
For instance, transitioning from LO to NLO calculations in the $pp \to \wpm jjj$ process 
results in about a 90\% reduction in scale uncertainty. 
Therefore, in this paper, we do not further consider scale uncertainties.\footnote{Concerns might arise 
regarding certain processes that maintain high uncertainty at NLO. 
One example is $pp \to \gamma\gamma j$, 
which at LO shows a cross section of $17.56 ^{+17.92\%}_{-15.77\%} \pb$, 
and at NLO, $41.11^{+16.15\%}_{-14.38\%}\pb$. 
This is based on the standard approach for $\mu_R$ and $\mu_F$ choices, 
considering the maximum value as $2\mu_0$ and the minimum as $\mu_0/2$. 
To determine the uncertainty in the $K$-factor, however,
it is reasonable to first fix the scale choice from nine options
and then compute the uncertainty. 
In this approach, we find $K(\gamma\gamma j) = 2.34^{+5.66\%}_{-3.54\%}$, 
showcasing significantly less uncertainty compared to the LO or NLO cross sections alone.}

\begin{table}[h!]
    \centering
 \begin{tabular}{@{\hspace{1em}}l@{\hspace{1em}}|@{\hspace{1em}}l|@{\hspace{1em}}c@{\hspace{1em}}|c|c}
      \toprule
    \multicolumn{5}{c}{Vector boson (+ jets) at the 14 TeV LHC} \\  \hline 
    Process & Syntax & $\sg_{\rm LO}~[{\rm pb}]$ & ~$K$-factor~  & ~Ref.~ \\
    \hline
$pp\to \wpm$ & \texttt{p p >  wpm }     & $1.390 \times 10^{5}$ & $1.381$  & \\
$pp\to \wpm j$ & \texttt{p p > wpm j}      & $2.289\times 10^{4}$ & $1.401$ & \cite{Campbell:2003hd}  \\
$pp\to \wpm jj$ &  \texttt{p p > wpm j j}      & $7.398\times 10^3$ & 1.234 &\cite{Campbell:2003hd}  \\
$pp\to \wpm jjj$ &  \texttt{p p > wpm j j j }    & $2.055 \times 10^{3}$ & $1.115$  & \\
    \hline
$pp\to Z$ & \texttt{p p > z }   & $4.213\times 10^{4}$ & $1.371$  &  \\
$pp\to Zj$ & \texttt{p p > z j }   & $7.531  \times 10^{3}$ & $1.381$  & \\
$pp\to Zjj$ & \texttt{p p > z j j } & $2.387 \times 10^{3}$ & $1.227$  &  \\
$pp\to Zjjj$ & \texttt{p p > z j j j } & $6.671 \times 10^{2}$ & $1.113$  &  \\
    \hline
$pp\to \gm j$ & \texttt{p p > a j }  & $2.602  \times 10^{4}$ & $2.952$  &\\
$pp\to \gm jj$ & \texttt{p p > a j j}   & $1.175  \times 10^{4}$ & $1.466$   &\\
    \hline
    \end{tabular} 
     \caption{$K$-factors for single vector boson production and associated production with jets. The label {\tt wpm} encompasses both $W^+$ and $W^-$, as specified in the command with {\tt define wpm = w+ w-}.
}
     \label{tab:V:jet}
\end{table}

Let us embark on an extensive exploration of the $K$-factors, 
beginning with the production of single electroweak gauge vector bosons 
($\wpm$, $Z$, and $\gamma$) and their associated production with jets.
For the sake of brevity, these will be collectively referred to as vector bosons. 
In Table \ref{tab:V:jet}, 
we report the LO total cross section in picobarns (pb) along with their corresponding $K$-factors, as defined in \eq{eq:K:factor}. 
Additionally, we specify the syntax for each process, 
indicating  that  {\tt wpm} collectively represents $W^+$ and $W^-$
 (defined in the command shell as {\tt define wpm = w+ w-}).   
The $K$-factors display significant variation across different processes, 
highlighting the diverse impact of NLO corrections.

For processes $pp \to V$ without jets, 
the variation in the $K$-factor, $\delta K$, is approximately 40\%. 
Adding a jet to the single $V$ production process
yields a slight increase in $\delta K$, about 1\% higher than what is observed for the $pp \to V$ process alone.
However, incorporating more jets leads to a noticeable reduction in $\delta K$. Specifically, when the single $V$ production includes three jets, $\delta K$ reduces to around 11\%, showcasing a trend where $\delta K$ decreases with the inclusion of each additional jet.

An especially noteworthy observation from Table \ref{tab:V:jet} is the significantly  high $K$-factors 
for photon productions accompanied by a jet, reaching approximately three. 
This pattern of elevated $K$-factors in photon-involved productions is consistent across various processes, a phenomenon attributed to the different production mechanisms at LO and NLO.
At LO, photon production predominantly occurs through quark-antiquark scattering, 
while NLO introduces the possibility of gluon-initiated processes. 
With gluon Parton Distribution Functions (PDFs) markedly increasing with $Q^2$, gluon-initiated processes become dominant at the LHC's high-energy scales once they are allowed.

\begin{table}[h!]
    \centering
 \begin{tabular}{@{\hspace{1em}}l|@{\hspace{1em}}l|@{\hspace{1em}}c@{\hspace{1em}}|c|c|c}
      \toprule
    \multicolumn{6}{c}{Two vector Bosons (+ jets) at the 14 TeV LHC} \\  \hline 
    Process & Syntax & $\sg_{\rm LO}~[{\rm pb}]$ & ~$K$-factor~ & Note &  ~Ref.~ \\
    \hline
$pp\to \ww$ & \texttt{p p > w+ w- }   & $8.213  \times 10$ & $1.417$ &   & \cite{Campbell:1999ah}\\
$pp\to ZZ$ & \texttt{p p > z z }   & $1.179  \times 10$ & $1.316$ & $K_{\rm NNLO}=1.72$\cite{Cascioli:2014yka} &  \cite{Campbell:1999ah,Cascioli:2014yka}\\
$pp\to Z \wpm$ & \texttt{p p > z wpm }   & $3.158  \times 10$ & $1.599$ &   & \cite{Campbell:1999ah}\\
$pp\to \rr$ & \texttt{p p > a a }   & $3.856  \times 10$ & $2.777$ &   & \cite{Campbell:1999ah}\\
$pp\to \gm Z$ & \texttt{p p > a z }   & $3.346 \times 10$ & $1.498$ &   & \cite{Campbell:1999ah}\\
$pp\to \gm \wpm$ & \texttt{p p > a wpm }   & $3.744\times 10$ & $2.667$ &   & \cite{Campbell:1999ah}\\
    \hline
$pp\to \ww j$ & \texttt{p p > w+ w- j}   & $3.332  \times 10$ & $1.322$ &   & \\
$pp\to ZZ j$ & \texttt{p p > z z j}   & $3.938 $ & $1.367$ &   & \\
$pp\to Z \wpm j$ & \texttt{p p > z wpm j}   & $1.854\times 10 $ & $1.304$ &   & \\
$pp\to \rr j$ & \texttt{p p > a a j}   & $1.761\times 10 $ & $2.528$ &   & \\
$pp\to \gm Z j$ & \texttt{p p > a z j}   & $1.197\times 10 $ & $1.571$ &   & \\
$pp\to \gm \wpm j$ & \texttt{p p > a wpm j}   & $3.501\times 10 $ & $1.594$ &   & \\
    \hline
$pp\to W^+ W^+ jj$ & \texttt{p p > w+ w+ j j}   & $1.619 \times 10^{-1}$ & $1.610$ &   & \\
$pp\to W^- W^- jj$ & \texttt{p p > w- w- j j}   & $7.064 \times 10^{-2}$ & $1.663$ &   & \\
$pp\to W^+ W^- jj$ & \texttt{p p > w+ w- j j}   & $1.388\times 10$ & $1.202$ &   & \\
$pp\to ZZ jj$ & \texttt{p p > z z j j}   & $1.437 $ & $1.302$ &   & \\
$pp\to Z \wpm j j$ & \texttt{p p > z wpm j j}   & $9.223 $ & $1.149$ &   & \\
$pp\to \rr jj$ & \texttt{p p > a a j j}   & $1.164\times 10 $ & $1.539$ &   & \\
$pp\to \gm Z jj$ & \texttt{p p > a z j j}   & $5.048 $ & $1.387$ &   & \\
$pp\to \gm \wpm jj$ & \texttt{p p > a wpm j j}   & $1.789\times 10 $ & $1.292$ &   & \\
 \hline
    \end{tabular} 
     \caption{$K$-factors for the production of two vector bosons, including processes with jets.
 }
     \label{tab:VV:jet}
\end{table}

In Table \ref{tab:VV:jet}, our focus shifts to the production of double vector bosons, 
incorporating scenarios both without and with jet involvement. 
We present the LO cross sections in pb alongside their respective $K$-factors. 
Notably, we cite the NNLO $K$-factor for $ZZ$ production, 
$K_{\text{NNLO}}(ZZ) \approx 1.72$\cite{Cascioli:2014yka}, emphasizing the importance of higher-order corrections for accurate predictions.

Our analysis indicates that photon-involved processes command exceptionally high $K$-factors,
as in the cases of single vector boson production.
In particular,  $\gamma\gamma$ and $\gamma W^\pm$ processes
stand out with $K$-factors of approximately $2.78$ and $2.67$, respectively. 
Table \ref{tab:VV:jet} also demonstrates that including jets in double vector boson productions leads to a decrease in $\delta K$, suggesting an anti-correlation between the number of jets and $\dt K$.

A key finding involves the same-sign $W$ boson pair production processes, 
$W^+ W^+ jj$ and $W^- W^- jj$, 
uniquely occurring with two jets. 
Despite their relatively low total cross-sections,
$\sigma_{\text{LO}}(pp \to W^+ W^+ jj) \approx 162 \fb$ and $\sigma_{\text{LO}}(pp \to W^- W^- jj) \approx 71 \fb$,
these processes are crucial as backgrounds in searches for BSM signals leading to same-sign dilepton final states.
The notable $K$-factors, $K(W^+ W^+ jj) \approx 1.61$ and $K(W^- W^- jj) \approx 1.66$, underscore their impact in BSM signal-to-background analyses.

In addition, our research delves into gluon fusion production of $\gamma\gamma$, $W^+W^-$, $ZZ$, and $Z\gamma$,
a topic not covered in Table \ref{tab:VV:jet} but essential for a comprehensive understanding of both SM and BSM physics. 
Since the table presents LO and NLO cross-sections within perturbative calculations, 
gluon fusion is not included. 
The LO processes proceed through quark-antiquark annihilation ($q\bar{q} \to VV'$) with $\alpha_{\text{em}}^2$ coupling. 
At NLO in QCD, virtual corrections and real emissions introduce $\alpha_s\alpha_{\text{em}}^2$, 
occurring via quark-antiquark annihilation and gluon-quark scattering. 
Gluon fusion channels, enabled by quark box diagrams, exhibit $\alpha_s^2\alpha_{\text{em}}^2$ coupling and are thus considered at the NNLO level.
However, the markedly higher gluon PDF, compared to the quark PDF, compensates for the extra $\alpha_s$ factor, highlighting the importance of these processes in detailed particle physics analyses.

Our calculations of the gluon fusion production cross-sections at the 14 TeV LHC are as follows:
\begin{align}
\sigma(gg \to \gamma\gamma) &= 3.451 \times 10 \pb, \\ \nn
\sigma(gg \to W^+W^-) &= 4.412 \pb,  \\ \nn
\sigma(gg \to ZZ) &= 1.450 \pb,  \\ \nn
\sigma(gg \to Z\gamma) &= 8.827 \times 10^{-1}  \pb.
\end{align}
Although these cross-sections do not encompass the complete NNLO results,
they provide significant insights. 
For photon pair production, the gluon fusion process is as critical as the LO quark-antiquark annihilation.
Incorporating this gluon fusion process, the $K$-factor escalates to 3.67. 
For the other three vector boson production processes, however, contributions from gluon fusion are relatively minor, with the cross-sections constituting only a few percent of the LO cross-section.

\begin{table}[h!]
    \centering
 \begin{tabular}{@{\hspace{1em}}l @{\hspace{1em}} |@{\hspace{1em}}l@{\hspace{1em}}|@{\hspace{1em}}c@{\hspace{1em}}|c|c}
      \toprule
    \multicolumn{5}{c}{Three Vector Bosons (+ jets) at the 14 TeV LHC} \\  \hline 
    Process & Syntax & $\sg_{\rm LO}~[{\rm pb}]$ & ~$K$-factor~  & ~Ref.~ \\
    \hline
$pp\to \ww \wpm$ & \texttt{p p > w+ w- wpm}   & $1.542  \times 10^{-1}$ & $1.565$ &   \cite{Binoth:2008kt,Dittmaier:2017bnh}\\
$pp\to Z\ww$ & \texttt{p p > z w+ w- }   & $1.172  \times 10^{-1}$ & $1.659$ &   \cite{Binoth:2008kt}\\
$pp\to Z Z \wpm$ & \texttt{p p > z z wpm }   & $3.724\times 10^{-2}$ & $1.763$   & \cite{Binoth:2008kt}\\
$pp\to Z Z Z$ & \texttt{p p > z z z }   & $1.239\times 10^{-2}$ & $1.320$ &    \cite{Binoth:2008kt}\\
$pp\to \gm \ww$ & \texttt{p p > a w+ w- }   & $1.933 \times 10^{-1}$ & $1.830$ &    \\
$pp\to \rr \wpm$ & \texttt{p p > a a wpm }   & $3.956\times 10^{-2}$ & $3.607$ &    \cite{CMS:2021jji}\\
$pp\to \gm Z \wpm$ & \texttt{p p > a z wpm }   & $6.838\times 10^{-2}$ & $2.274$ &    \\
$pp\to \gm Z Z$ & \texttt{p p > a z z }   & $3.132\times 10^{-2}$ & $1.356$ &    \\
$pp\to \rr Z $ & \texttt{p p > a a z }   & $4.949\times 10^{-2}$ & $1.569$ &   \cite{CMS:2021jji} \\
$pp\to \rr \gm $ & \texttt{p p > a a a }   & $2.403\times 10^{-2}$ & $2.995$ &    \\
    \hline
$pp\to \ww \wpm j $ & \texttt{p p > w+ w- wpm j}   & $1.109 \times 10^{-1}$ & $1.307$ &   \\
$pp\to Z\ww j$ & \texttt{p p > z w+ w- j}   & $9.957 \times 10^{-2}$ & $1.278$ &    \\
$pp\to Z Z \wpm j$ & \texttt{p p > z z wpm j}   & $3.348\times 10^{-2}$ & $1.299$ &    \\
$pp\to Z Z Z j$ & \texttt{p p > z z z j}   & $5.579\times 10^{-3}$ & $1.343$ &    \\
$pp\to \rr \wpm j$ & \texttt{p p > a a wpm j}   & $7.073\times 10^{-2}$ & $1.623$ &    \\
$pp\to \gm Z \wpm j$ & \texttt{p p > a z wpm j}   & $8.356\times 10^{-2}$ & $1.405$    & \\
$pp\to \gm Z Z j$ & \texttt{p p > a z z j}   & $1.426\times 10^{-2}$ & $1.410$ &    \\
$pp\to \rr Z j$ & \texttt{p p > a a z j}   & $2.490\times 10^{-2}$ & $1.596$ &   \\
$pp\to \rr \gm j$ & \texttt{p p > a a a j}   & $2.312\times 10^{-2}$ & $2.332$ &   \\
 \hline
    \end{tabular} 
     \caption{The $K$-factors for three  vector boson production, including the cases with jets. }
     \label{tab:VVV:jet}
\end{table}

\begin{table}[h!]
    \centering
 \begin{tabular}{@{\hspace{1em}}l @{\hspace{1em}} |@{\hspace{1em}}l@{\hspace{1em}}|@{\hspace{1em}}c@{\hspace{1em}}|c|c}
      \toprule
    \multicolumn{5}{c}{Four Vector Bosons (+ jets) at the 14 TeV LHC} \\  \hline 
    Process & Syntax & $\sg_{\rm LO}~[{\rm pb}]$ & ~$K$-factor~  & ~Ref.~ \\
    \hline
$pp\to \ww \ww$ & \texttt{p p > w+ w- w+ w-}   & $7.235 \times 10^{-4}$ & $1.638$ &  \\
$pp\to \ww \wpm Z$ & \texttt{p p > w+ w- wpm z}   & $8.509 \times 10^{-4}$ & $1.729$ &   \\
$pp\to \ww \wpm \gm$ & \texttt{p p > w+ w- wpm a}   & $1.141 \times 10^{-3}$ & $1.856$ &   \\
$pp\to \ww ZZ$ & \texttt{p p > w+ w- z z}   & $5.483 \times 10^{-4}$ & $1.554$ &  \\
$pp\to \ww Z\gm$ & \texttt{p p > w+ w- z a}   & $1.176\times 10^{-3}$ & $1.733$ &   \\
$pp\to \ww \rr$ & \texttt{p p > w+ w- a a}   & $8.291\times 10^{-4}$ & $1.869 $ &  \\
$pp\to \wpm ZZZ $ & \texttt{p p > wpm z z z}   & $7.713\times 10^{-5}$ & $1.967 $ &   \\
$pp\to \wpm ZZ\gm  $ & \texttt{p p > wpm z z a}   & $1.677\times 10^{-4}$ & $2.526 $ &   \\
$pp\to \wpm Z\rr  $ & \texttt{p p > wpm z a a}   & $1.760\times 10^{-4}$ & $3.089$ &  \\
$pp\to \wpm \gm \rr  $ & \texttt{p p > wpm a a a}   & $6.719\times 10^{-5}$ & $4.221$ &   \\
$pp\to ZZZZ  $ & \texttt{p p > z z z z}   & $2.449\times 10^{-5}$ & $1.262$ &   \\
$pp\to ZZZ\gm  $ & \texttt{p p > z z z a}   & $5.607\times 10^{-5}$ & $1.270 $ &  \\
$pp\to ZZ\rr  $ & \texttt{p p > z z a a}   & $9.284\times 10^{-5}$ & $1.363 $ &   \\
$pp\to Z\gm\rr  $ & \texttt{p p > z a a a}   & $9.716\times 10^{-5}$ & $1.585$ &  \\
$pp\to \rr\rr  $ & \texttt{p p > a a a a}   & $3.964\times 10^{-5}$ & $2.405$ &  \\
 \hline
    \end{tabular} 
     \caption{The $K$-factors for four vector boson production, including the cases with jets. }
     \label{tab:4V:jet}
\end{table}

Moving to more intricate scenarios, we analyze productions involving three or four vector bosons, 
with and without jet accompaniment, in Table \ref{tab:VVV:jet} and Table \ref{tab:4V:jet}. 
Here, we report the LO cross sections and their corresponding $K$-factors, citing key studies that provide detailed calculations or analyses in both theoretical and experimental contexts.
It is consistently observed that photon-involved processes exhibit notably high $K$-factors, 
such as $K(\gamma\gamma W^\pm) \simeq 3.6$, $K(\gamma\gamma\gamma) \simeq 3.0$, and $K(W^\pm\gamma\gamma\gamma) \simeq 4.2$.

Table \ref{tab:VVV:jet} also illustrates the effect of jet inclusion on $K$-factors. 
Similar to single and double vector boson productions,
adding jets tends to reduce $\delta K$. 
This reduction is particularly pronounced in scenarios where productions of three or four vector bosons without jets initially exhibit high $K$-factors.
To illustrate, consider two scenarios: 
$W^+ W^-W^\pm$ with a conventional $K$-factor, 
versus $\gamma\gamma W^\pm$, which has a notably high $K$-factor. 
The inclusion of an additional jet presents contrasting impacts: 
for $W^+ W^-W^\pm$, the $K$-factor of approximately 1.57 is reduced by about 17\%, 
whereas for $\gamma\gamma W^\pm$, with a $K$-factor of approximately 3.61, the reduction is around 55\%.

\begin{table}[!ht]
    \centering
 \begin{tabular}{@{\hspace{0.5em}}l|@{\hspace{0.5em}}l|@{\hspace{0.5em}}c@{\hspace{0.5em}}|c|c|c}
      \toprule
    \multicolumn{6}{c}{Light jet or Bottom quark productions at the 14 TeV LHC} \\  \hline 
    Process & Syntax & $\sg_{\rm LO}~[{\rm pb}]$ & $K$-factor & Note & ~Ref.~ \\
    \hline
    $pp\to jj$ & \texttt{p p >  j  j}     & $1.270 \times 10^{6}$ & $1.290$ & \multirow{2}{*}{$\pt^{j}>100\gev$} &     \\
$pp\to jjj$ & \texttt{p p >  j j j}    & $3.431 \times 10^{4}$ & $1.005$ &  &     \\
\hline
$pp\to b\bar{b}$ & \texttt{p p >  b b$\sim$}    & $2.972\times 10^{8}$ & $1.348$ & $K_{\rm NNLO}\simeq 1.30$\cite{Catani:2020kkl} & \\ \hline
$pp\to b\bar{b}j$ & \texttt{p p >  b b$\sim$ j }     & $4.780\times 10^{6}$ & $1.268$ &   &    \\
$pp\to b\bar{b}jj$ & \texttt{p p >  b b$\sim$ j j}     & $8.903\times 10^{5}$ & $1.499$ &   \textsc{MG}
\texttt{v2.9.16} &    \\
$pp\to b\bar{b}b\bar{b}$ & \texttt{p p >  b b$\sim$ b b$\sim$}     & $3.714\times 10^{5}$ & $1.997$ &   \textsc{MG}
\texttt{v2.9.16} &    \\
   \hline
$pp\to  b\bar{b} \wpm$ & \texttt{p p >  b b$\sim$  wpm }     & $3.451\times 10^{2}$ & $2.808$ &  &  \cite{Badger:2010mg}  \\
$pp\to  b\bar{b}Z$ & \texttt{p p >  b b$\sim$ z }     & $8.493\times 10^{2}$ & $1.706$ &  &  \cite{Frederix:2011qg} \\
$pp\to b\bar{b} \gm $ & \texttt{p p >  b b$\sim$ a }     & $2.620\times 10^{3}$ & $2.182$ &  &  \cite{Frederix:2011qg} \\
$pp\to  b\bar{b}\wpm j$ & \texttt{p p > b b$\sim$   wpm j}     & $2.246\times 10^{2}$ & $2.231$ &  &   \\
$pp\to b\bar{b}  Zj$ & \texttt{p p > b b$\sim$  z j}     & $1.891\times 10^{2}$ & $1.896$ & MG v3.3.1 &   \\
$pp\to b\bar{b}  \gm j$ & \texttt{p p > b b$\sim$  a j}     & $1.133\times 10^{3}$ & $1.763$ &  &   \\
   \hline
    \end{tabular} 
     \caption{The $K$-factors for light jets or bottom quark production. 
We do not put any cut on $b$ and $\bar{b}$ quarks.
}
     \label{tab:j:b}
\end{table}

\begin{table}[!ht]
    \centering
 \begin{tabular}{@{\hspace{0.5em}}l|@{\hspace{0.5em}}l|@{\hspace{0.5em}}c@{\hspace{0.5em}}|c|c|c}
      \toprule
    \multicolumn{6}{c}{Top quark productions at the 14 TeV LHC} \\  \hline 
    Process & Syntax & $\sg_{\rm LO}~[{\rm pb}]$ & $K$-factor & Note & ~Ref.~ \\
    \hline
    $pp\to t \bar{b}/\bar{t}b$ & \texttt{p p > t b$\sim$ \& p p >  t$\sim$ b }   & $8.495$ & $1.348$ &  & \\
$pp\to \ttop$ & \texttt{p p > t t$\sim$}   & $5.302\times 10^{2}$ & $1.489$ & $K_{\rm aN^3LO}=1.719$\cite{Kidonakis:2022hfa}& \cite{Nason:1987xz,Beenakker:1990maa,Czakon:2011xx,Kidonakis:2019yji} \\
$pp\to \ttop j$ & \texttt{p p > t t$\sim$ j }   & $3.522  \times 10^{2}$ & $1.422$ &  & \\
$pp\to \ttop jj$ & \texttt{p p > t t$\sim$ j j}   & $1.550 \times 10^{2}$ & $1.426$ &  & \\
$pp\to \ttop \ttop$ & \texttt{p p > t t$\sim$ t t$\sim$}   & $5.135 \times 10^{-3}$ & $2.284$ &  & \cite{Bevilacqua:2012em,ATLAS:2023ajo,CMS:2023ftu}\\
$pp\to \ttop b \bar{b}$ & \texttt{p p > t t$\sim$ b b$\sim$}   & $7.500$ & $2.513$ & MG v3.3.2 &  \cite{Bredenstein:2009aj} \\
   \hline
$pp\to  \ttop \wpm$ & \texttt{p p > t t$\sim$ wpm }   & $4.685 \times 10^{-1}$ & $1.456$ &  & \\
$pp\to  \ttop Z $ & \texttt{p p > t t$\sim$ z }   & $6.103 \times 10^{-1}$ & $1.511$ &  & \\
$pp\to\ttop  \gm $ & \texttt{p p > t t$\sim$ a }   & $1.513$ & $1.560$ &  & \\
   \hline
$pp\to \ttop  \wpm j$ & \texttt{p p > t t$\sim$ wpm j}   & $2.898 \times 10^{-1}$ & $1.496$ &  & \\
 $pp\to \ttop  Z j$ & \texttt{p p > t t$\sim$ z j}   & $4.512 \times 10^{-1}$ & $1.424$ &  & \\
$pp\to  \ttop \gm j$ & \texttt{p p > t t$\sim$ a j }   & $1.134$ & $1.456$ &  & \\
   \hline
$pp\to \ttop  \ww$ & \texttt{p p > t t$\sim$ w+ w- }   & $8.492\times 10^{-3}$ & $1.466$ &  & \\
$pp\to  \ttop \wpm Z$ & \texttt{p p > t t$\sim$ wpm z }   & $3.189\times 10^{-3}$ & $1.389$ &  & \\
$pp\to \ttop \wpm \gm $ & \texttt{p p > t t$\sim$ wpm a }   & $3.933\times 10^{-3}$ & $1.439$ &  & \\
$pp\to  \ttop ZZ$ & \texttt{p p > t t$\sim$ z z }   & $1.665\times 10^{-3}$ & $1.380$ &  & \\
$pp\to  \ttop Z\gm$ & \texttt{p p > t t$\sim$ z a }   & $3.318\times 10^{-3}$ & $1.531$ &  & \\
$pp\to \ttop  \rr$ & \texttt{p p > t t$\sim$ a a }   & $4.924\times 10^{-3}$ & $1.471$ &  & \\
$pp\to \ttt \wpm $ & \texttt{p p > tt wpm }   & $5.462\times 10$ & $1.415$ & MG v2.9.16, 5 flavor& \\
      \hline
    \end{tabular} 
     \caption{The $K$-factors for top quark production. 
     Here $\ttt$ (in syntax, {\tt tt})  is a label that includes both $t$ and $\bar{t}$, 
defined from the shell with {\tt define tt = t t$\sim$}. Here $K_{\rm aN^3LO}$ denote the $K$-factor
for the approximate NNNLO. We do not put any cut on $b$ and $\bar{b}$ quarks.
}
     \label{tab:b:t}
\end{table}

In Table \ref{tab:j:b} and Table \ref{tab:b:t}, we explore scenarios involving light jets,
bottom quarks, top quarks, and their association with vector bosons and/or jets. The LO cross sections and their respective $K$-factors are meticulously detailed. Here, $\ttt$ (in syntax, {\tt tt}) includes both $t$ and $\bar{t}$, 
established with the command {\tt define tt = t t$\sim$}. 
No kinematic cuts are applied to the bottom and top quarks in our analysis.

Additionally, we reference higher-level corrections for $\bb$ and $\ttop$: 
$K_{\rm NNLO}(\bb) = 1.25\sim1.35$\cite{Catani:2020kkl}  and 
$K_{\rm aN^3LO}(\ttop) = 1.719$\cite{Kidonakis:2022hfa}, 
where  $K_{\rm aN^3LO}$ denote the $K$-factor
for the approximate NNNLO.
We also point out the employment of different  {\sc\small MadGraph} software versions for particular processes. 
 Importantly,  {\sc\small MadGraph}  \texttt{v2.9.16} is used for $b\bar{b}jj$ and $b\bar{b}b\bar{b}$ processes 
because this version uniquely offers stable and reliable NLO results, 
an attribute not shared by higher versions.

Including a vector boson in bottom quark pair production processes significantly increases the $K$-factors.
This effect is highlighted by the $pp \to  b\bar{b}W^\pm$ process, 
with a $K$-factor of $2.808$, 
and the $pp \to  b\bar{b}\gamma j$ process, showcasing a $K$-factor of $1.763$. 
These elevated $K$-factors significantly impact the search for new particles, 
particularly those decaying mainly into bottom quark pairs. 
Examples of interest include new neutral Higgs bosons within Two-Higgs-Doublet Models
or the Minimal Supersymmetric Standard Model, 
as well as new $Z'$ bosons in extended gauge symmetry models. 
The challenge posed by dijet backgrounds in $b\bar{b}$ resonance searches 
often necessitates the associated production of an additional vector boson, 
underscoring the critical need to account for high $K$-factors in detailed background analyses.

\begin{table}[!ht]
    \centering
 \begin{tabular}{@{\hspace{1em}}l@{\hspace{1em}}|@{\hspace{1em}}l|@{\hspace{1em}}c@{\hspace{1em}}|c|@{\hspace{1em}}c}
      \toprule
    \multicolumn{5}{c}{Higgs associated processes at the 14 TeV LHC} \\  \hline 
    Process & Syntax & $\sg_{\rm LO}~[{\rm pb}]$ & $K$-factor & Note  \\
    \hline
$pp\to H \wpm$ & \texttt{p p >  h wpm}     & $1.352$ & $1.183$ &       \\
$pp\to H \wpm j$ & \texttt{p p >  h wpm j}     & $4.638\times 10^{-1}$ & $1.215$ &       \\
$pp\to H \wpm j j$ & \texttt{p p >  h wpm j j}     & $1.518\times 10^{-1}$ & $1.196$ &       \\
    \hline
$pp\to H Z$ & \texttt{p p >  h z}     & $7.176 \times 10^{-1}$ & $1.192$ &       \\
$pp\to H Zj$ & \texttt{p p >  h z j}     & $2.475\times 10^{-1}$ & $1.234$ &       \\
$pp\to H Zjj$ & \texttt{p p >  h z j j}     & $8.099\times 10^{-2}$ & $1.212$ &       \\
    \hline
$pp\to H \ww$ & \texttt{p p >  h w+ w-}     & $1.009\times 10^{-2}$ & $1.224$ &       \\
$pp\to H \wpm \gm$ & \texttt{p p >  h wpm a}     & $3.418\times 10^{-3}$ & $1.328$ &       \\
$pp\to H \wpm Z$ & \texttt{p p >  h wpm z}     & $4.592\times 10^{-3}$ & $1.336$ &       \\
$pp\to H Z Z$ & \texttt{p p >  h z z}     & $2.467\times 10^{-3}$ & $1.198$ &       \\
    \hline
$pp\to H \ttop$ & \texttt{p p >  h t t$\sim$}     & $4.168\times 10^{-1}$ & $1.332$ &       \\
 $pp\to H \ttt j$ & \texttt{p p >  h tt j}     & $4.773\times 10^{-2}$ & $1.535$ &  5-flavor     \\
$pp\to H b \bar{b}$ & \texttt{p p >  h b b$\sim$}     & $5.248\times 10^{-1}$ & $1.146$ &      \\
$pp\to H b \bar{b} j$ & \texttt{p p >  h b b$\sim$ j}     & $8.504\times 10^{-2}$ & $1.306$ & MG v3.3.1     \\
    \hline
$pp\to H H \wpm$ & \texttt{p p >  h h wpm}     & $5.250\times 10^{-4}$ & $1.107$ &       \\
$pp\to H H Z$ & \texttt{p p >  h h z}     & $3.248\times 10^{-4}$ & $1.071$ &       \\
    \hline
    \end{tabular} 
     \caption{The $K$-factors for the Higgs productions 
     associated with vector bosons, bottom quarks, or top quarks. 
Here $K_{\rm aN^3LO}$ denote the $K$-factor
for the approximate NNNLO. We do not put any cut on bottom and top quarks.
}
     \label{tab:H:jet}
\end{table}

Our analysis concludes in Table \ref{tab:H:jet} with an examination of the $K$-factors for various Higgs production processes 
at the 14 TeV LHC, covering associations with vector bosons, top quarks, bottom quarks, or multiple jets.  
This part highlights the multifaceted nature of Higgs production and the significance of investigating diverse channels to fully understand Higgs physics.  
The $K$-factors across these processes generally show modest increases, typically ranging from about $1.07$ to $1.5$.

We first focus on Higgs production associated with top or bottom quarks, instrumental 
in directly probing their Yukawa couplings. 
The $K$-factors for these processes are noteworthy: the $pp \to H \ttop$ process exhibits a $K$-factor of $1.332$, while the $pp \to H b \bar{b}$ process reaches a $K$-factor of $1.146$. 
These findings underscore the importance of accurate theoretical predictions in analyzing the interactions between the Higgs boson and heavy quarks. 
Another significant observation is in the $pp \to H\ttop j$ process, 
where the $K$-factor stands at $1.535$. 

Finally, we observe that double Higgs production processes associated with a vector boson, 
such as $pp \to HH \wpm$ and $pp \to HH Z$, 
have $K$-factors marginally above $1$. 
This subtle increment suggests that higher-order corrections 
for double Higgs production are less impactful than those for single Higgs production, 
marking a distinctive aspect of Higgs physics that merits thorough theoretical scrutiny.

\section{Panoramic view for high $K$-factor processes}
\label{sec:panoramic}

\begin{figure}[!t]
\centering
\includegraphics[width=\textwidth]{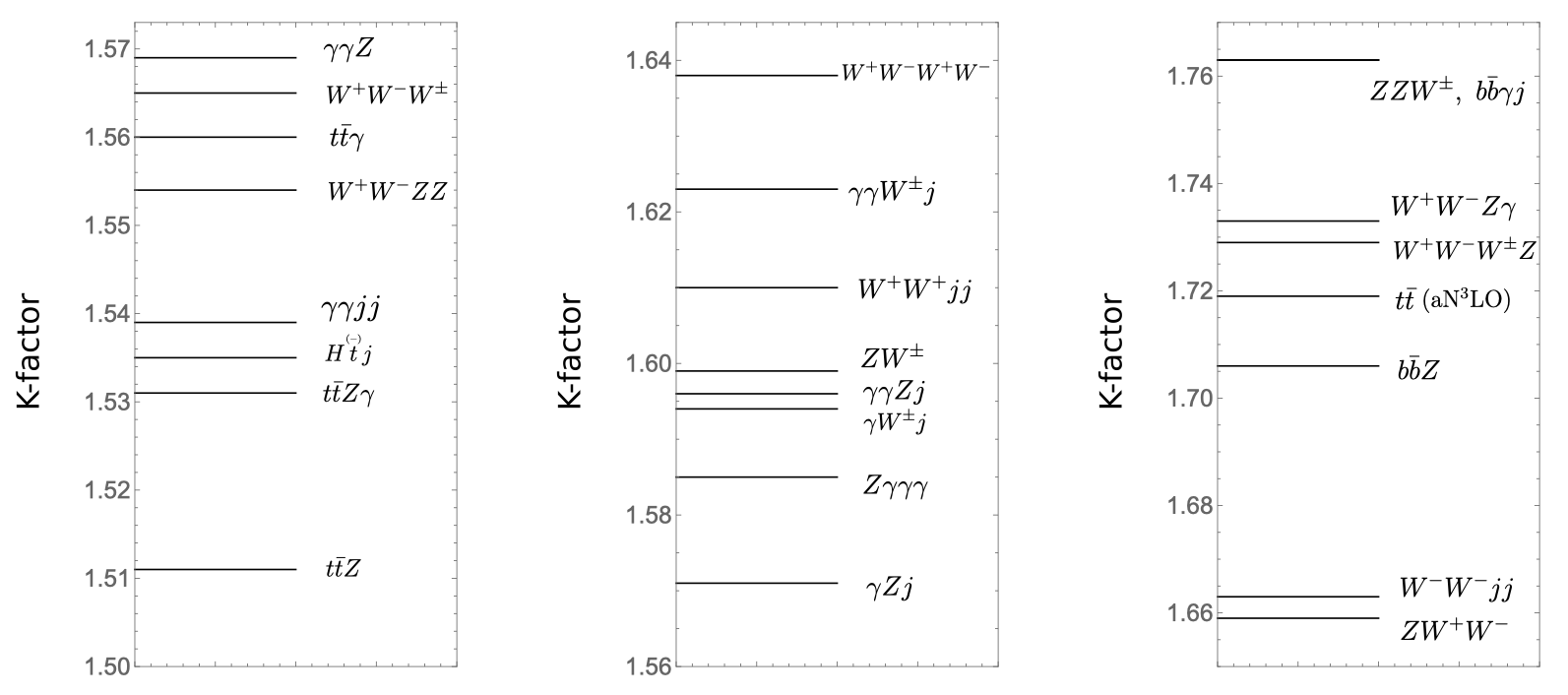}
\vspace{-0.7cm}
\caption{The processes with the $K$-factors in the range $[ 1.5,1.8 ]$. }
\label{fig-K1}
\end{figure}

In the preceding section, we conducted a thorough analysis of $K$-factors for 111 processes at the 14 TeV LHC. 
This section aims to synthesize these findings into a concise summary of processes that display high $K$-factors, offering an overview on processes where NLO corrections are notably significant.

Figure \ref{fig-K1} zeroes in on processes with $K$-factors ranging from $1.5$ to $1.8$. 
A prominent feature of these processes is the inclusion of multiple particles, especially vector bosons. Examples include $K(W^+ W^- W^\pm Z) = 1.729$ and $K(W^+W^-\gamma\gamma) = 1.869$. 
The elevated $K$-factors in these cases are attributed to the complexity of the interactions, 
which allow for multiple real emission processes.
Furthermore, $t\bar{t}$ production and their associated processes with $Z$, $\gamma$, or both, 
have notably high $K$-factors: 
$K_{\rm aN^3LO}(\ttop) \approx 1.72$, $K(\ttop Z) \approx 1.51$, $K(\ttop\gamma) \approx 1.56$, and $K(\ttop Z \gamma) \approx 1.53$. 
These figures highlight the indispensable role of higher-order corrections in top quark physics.

\begin{figure}[!t]
\centering
\includegraphics[width=\textwidth]{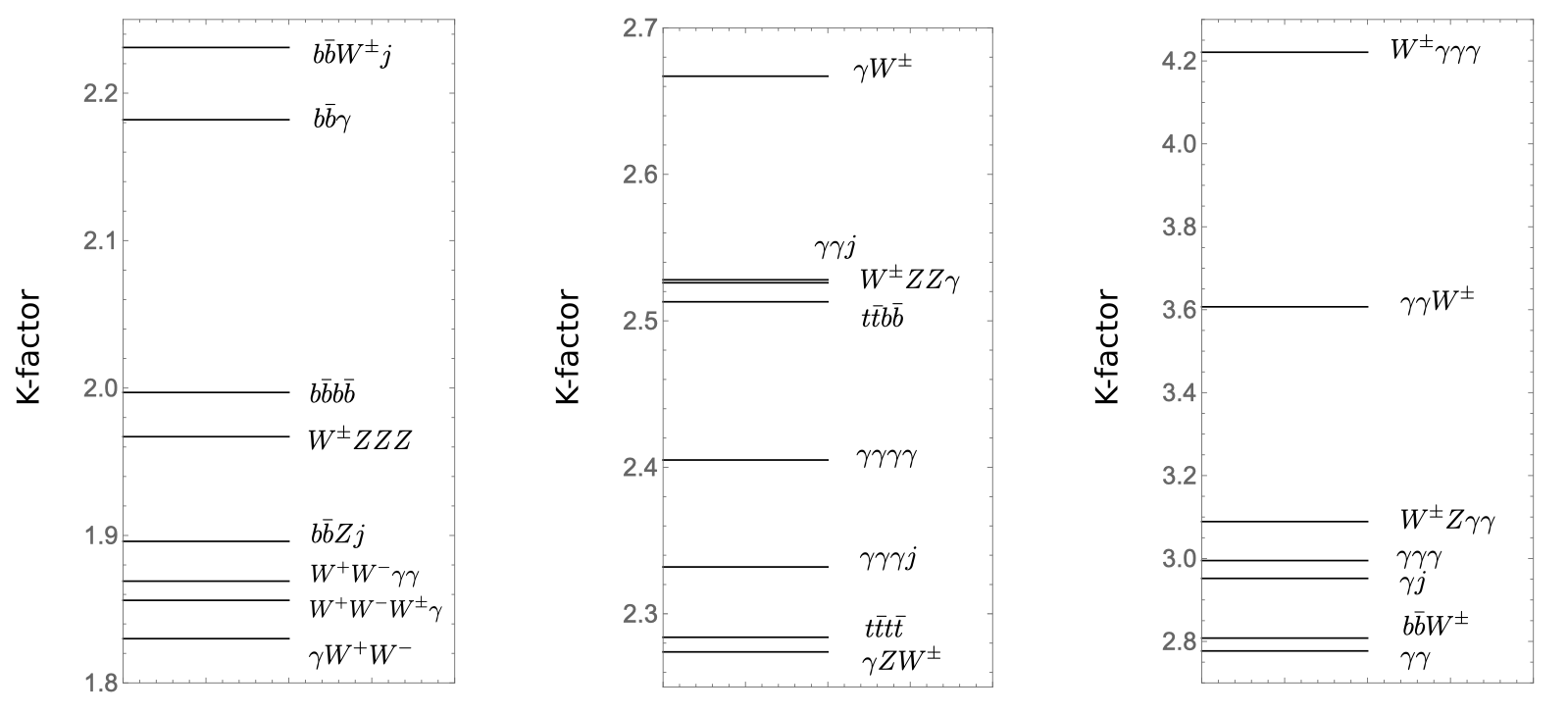}
\vspace{-0.7cm}
\caption{The processes with the $K$-factors larger than 1.8. }
\label{fig-K2}
\end{figure}

In Fig.~\ref{fig-K2}, we explore processes exhibiting $K$-factors above 1.8. 
 It is observed that the associated production of a bottom quark pair with various particles, such as $Z j$, $b\bar{b}$, $W^\pm j$, and $t\bar{t}$, all display $K$-factors exceeding 1.9. This observation is particularly relevant for BSM theories predicting new bosons that predominantly decay into a bottom quark pair, highlighting the importance of accounting for these high $K$-factors in background analyses for BSM searches.
 
Our second standout observation concerns the $K$-factor for the process of four top quark pair production ($t\bar{t}t\bar{t}$), which registers at 2.284. 
This process holds particular significance for exploring potential new physics scenarios, 
including a top-philic Axion-Like Particle (ALP)\cite{Blasi:2023hvb}
and two-Higgs-doublet model\cite{Anisha:2023vvu}.

The most significant trend from our comprehensive study on $K$-factors is the consistently high $K$-factors associated with photon-inclusive processes. Moreover, the $K$-factor values escalate with the addition of photons in the process. For instance, single $\wpm$ production associated with a photon
($\wpm \gamma $) has a $K$-factor of 2.667, which climbs to 4.221 for $W^\pm\gamma\gamma\gamma$.

The high $K$-factors for photon-inclusive processes are attributed to the introduction of a gluon as an initial particle at NLO, which is absent at LO. 
Consider  $pp \to \gamma W^+$.
This  process primarily occurs through $u\bar{d} \to \gamma W^+$ at LO.\footnote{For the sake of simplicity in our discussions, contributions from $c\bar{s} \to \gamma W^+$, while not negligible, are not mentioned.}
At NLO, there are three kinds of contributions:
(i) virtual corrections to $u\bar{d}  \to \gamma W^+$;
(ii) real emissions via $u\bar{d} \to \gamma W^+ g$;
(iii) real emissions through $g u \to \gamma W^+ d$ and $g\bar{d}   \to \gamma W^+ \bar{u}$, collectively represented as $gq \to W^+ \gamma q'$.

The $K$-factor for $\gamma W^+$ is broken down into:
\begin{align}
K ( \gm W^+)&= 
\frac{\sg_{\rm LO+virt}(u\bar{d} \to \gm W^+)}{\sg_{\rm LO}} + \frac{\sg_{\rm NLO}(u\bar{d}\to \gm W^+ g) }{\sg_{\rm LO}} + \frac{\sg_{\rm NLO} (g q \to W^+ \gm q')}{\sg_{\rm LO}}
\\ \nn
&\equiv K_{\rm virt} + K_{u\bar{d}\to \gm W^+ g} + K_{g q \to W^+ \gm q'}.
\end{align}
With the global $K$-factor of $K( \gm W^+) = 2.667$, contributions are as follows:
\begin{align}
\frac{K_{\rm virt} }{ K ( \gm W^+)} \approx 50.0\%,
\quad
\frac{K_{u\bar{d}\to \gm W^+ g} }{ K ( \gm W^+)} \approx  8.8\%,
\quad 
\frac{K_{g q \to W^+ \gm q'} }{ K ( \gm W^+)} \approx  41.2\%.
\end{align}
This breakdown clearly demonstrates how the high PDF of a gluon at the 14 TeV LHC 
enhances the real emission contribution from the initial gluon. 
Without this gluon contribution, the $K$-factor would be about 1.57, emphasizing the pivotal role of gluon-initiated processes in determining the $K$-factor.

\section{$K$-factors for differential distributions}
\label{sec:pt:invmass}

The search for new particles at the LHC heavily relies on the analysis of kinematic variables 
of detected particles.
Variables such as transverse momentum, invariant mass,
missing transverse energy, azimuthal angle differences, and rapidity gaps, 
are fundamental in probing BSM phenomena.
To effectively separate signals from backgrounds, 
analysis often concentrates on a limited parameter space, 
carefully selected based on the expected characteristics of new signals.
Consequently,  the distributions of $K$-factors across these kinematic variables emerge as a significant area of interest.

Two kinematic variables, transverse momentum and invariant mass, 
are particularly crucial in the search for new particles. 
The decay products of a heavy new particle typically manifest with elevated $p_T$, 
making high $p_T$ thresholds a strategic choice to reduce backgrounds from lower-energy SM activities. 
This strategy also aids in the effective triggering of signal events.
Equally critical is the invariant mass distribution of the decay products of a new particle, providing a straightforward method to determine the particle's mass.

Therefore, we analyze the $K$-factor distributions for both transverse momentum and invariant mass. To achieve accurate differential cross-section calculations at both LO and NLO,  
we employ the LO+PS  (Leading Order plus Parton Shower) and NLO+PS (Next-to-Leading Order plus Parton Shower) settings as detailed in Ref.\cite{Alwall:2014hca}. 
The LO+PS method computes matrix elements with NLO perturbative accuracy, 
incorporating both tree-level and one-loop matrix elements, and then matches these to parton showers.
This technique ensures that observables are reconstructed from the output of the Monte Carlo simulation.
The NLO+PS configuration extends the LO+PS methodology 
by basing its computations on NLO, 
and integrates NLO matrix elements with parton showers following the MC@NLO formalism.

\begin{figure}[!t]
\centering
\includegraphics[width=\textwidth]{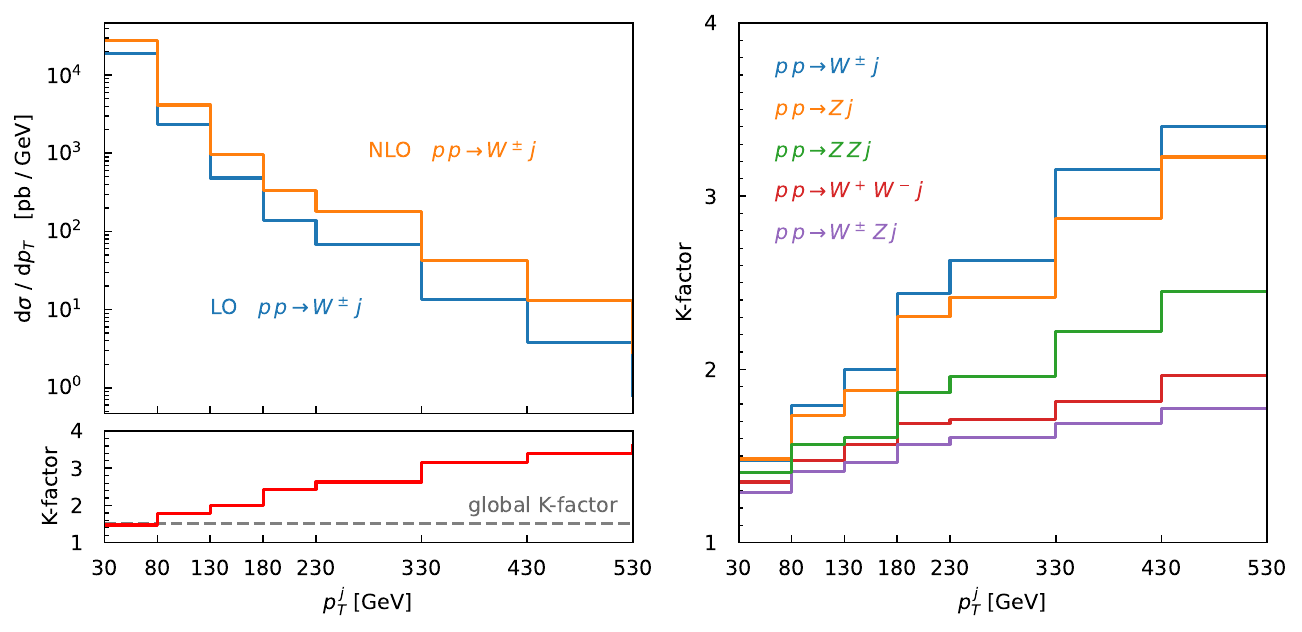}
\vspace{-0.7cm}
\caption{(Left-upper panel) Differential cross sections as a function of the transverse momentum of the jet, $p_T^j$, in the process $pp\to \wpm j$. The LO result is depicted by a blue solid line, while the NLO result is shown with an orange line.
(Left-lower panel) $K$-factor as a function of $p_T^j$. The horizontal black dashed line represents the $K$-factor for the total cross section.
(Right panel) $K$-factors as a function of $p_T^j$ for various processes: $pp\to \wpm j$ in blue, 
$pp\to Z j$ in orange, $pp\to ZZj$ in green, $pp\to \ww j$ in red, and $pp\to \wpm Z j$ in purple.}
\label{fig-K-ptj}
\end{figure}

For the transverse momentum dependence of $K$-factors, 
we analyze five processes: $pp\to \wpm j$, $pp\to Zj$, $pp\to ZZj$, $pp\to \ww j$, and $pp\to \wpm Z j$.\footnote{The $p_T^j$ dependence of $K$-factor was extensively studied in Ref.\cite{Rubin:2010xp}.}
These processes share similar global $K$-factor values, as outlined below:
\begin{align}
\label{eq:K:pt:processes}
K(\wpm j) \simeq K(Zj) \simeq K(ZZj) \simeq 1.4,\quad
K(\ww j) \simeq K(\wpm Z j) \simeq 1.3.
\end{align}

In Fig.~\ref{fig-K-ptj}, we present the $K$-factor distributions 
as a function of the jet's transverse momentum ($p_T^j$), 
by using {\sc\small Pythia} \texttt{v8.2}\cite{Sjostrand:2014zea}. 
The left panels specifically target the $pp\to \wpm j$ process. 
The upper left panel contrasts the differential cross sections 
at LO (depicted in blue) and NLO (shown in orange) with respect to $p_T^j$. 
We generated additional events at higher transverse momentum bins that suffered from low event counts.\footnote{Event counts at the $p_T$ threshold were adjusted to align with the differential cross sections.} 

The discrepancy between LO and NLO differential cross sections
becomes more pronounced  in high $p_T^j$ bins.
The lower left panel in Fig.~\ref{fig-K-ptj} details the differential $K$-factor in relation to $p_T^j$, 
where the global $K$-factor is indicated by a horizontal dashed black line. 
As the bulk of event counts falls within the lower $p_T^j$ bins, 
the global $K$-factor aligns with the $K$-factor in the $p_T^j$ bin of $[30,80]\gev$. 
Notably, the $K$-factor escalates as $p_T^j$ increases, surpassing three for $p_T^j>330\gev$. 
This underlines the critical need for careful background analysis 
 in high $p_T^j$ regions, 
when searching for a BSM signal where $\wpm j$ constitutes a primary background.

This pattern of a rising $K$-factor with increasing $p_T^j$ 
persists across the other four processes, 
as illustrated in the right panel of Fig.~\ref{fig-K-ptj}. 
Displayed are the $K$-factors for $pp\to \wpm j$ (blue), $pp\to Z j$ (orange), $pp\to ZZj$ (green), 
$pp\to \ww j$ (red), and $pp\to \wpm Z j$ (purple). 
While this trend is universally observed, 
the magnitude of increase varies, 
despite the processes sharing similar global $K$-factors, as noted in \eq{eq:K:pt:processes}. 
Remarkably, $pp\to \wpm j$ experiences the most significant rise, 
followed in sequence by $pp\to Z j$, $pp\to ZZj$, $pp\to \ww j$, and finally, $pp\to \wpm Z j$. 
At a pivotal $p_T^j$ of approximately 200\gev,
$pp\to \wpm j$ and $pp\to Z j$ report $K$-factors surpassing two, 
whereas the remaining processes display more moderate $K$-factor values.

\begin{figure}[!t]
\centering
\includegraphics[width=\textwidth]{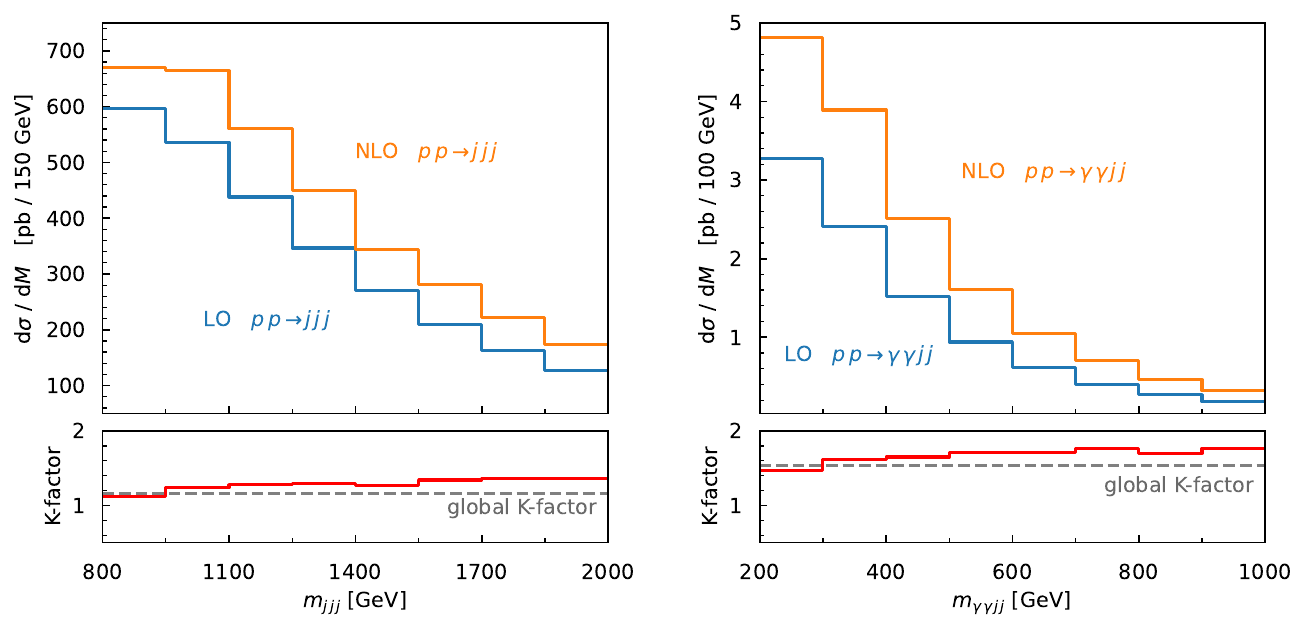}
\vspace{-0.7cm}
\caption{Differential cross sections about the invariant mass of three jets 
in the process of $pp\to jjj$ (left) and the invariant mass of $\rr jj$ in the process of $pp\to \rr jj$ (right).
In the upper panels, we present the LO results in blue and the NLO results in orange.
In the lower panels,
we present $K$-factors about the invariant mass.
The horizontal black dashed lines
denote the global $K$-factor for the total cross section. } 
\label{fig-K-mass}
\end{figure}

Shifting our focus to $K$-factor distributions related to invariant mass, we investigate two processes: $pp \to jjj$ and $pp \to \gamma\gamma jj$. The recent CMS Collaboration's search for narrow trijet resonances\cite{CMS:2023tep} emphasizes the relevance of the trijet process analysis. This research explores potential new particles, including a right-handed $Z$ boson decaying into three gluons\cite{Huitu:1996su}, a Kaluza–Klein gluon excitation decaying through an intermediate radion to three gluons\cite{Agashe:2016kfr,Agashe:2020wph}, and an excited quark decaying via a new boson\cite{Baur:1989kv}. In these analyses, the SM background is estimated using empirical functions to fit the $m_{jjj}$ spectrum, a technique challenging for phenomenological studies to mimic. Therefore, assessing whether the $K$-factor distribution for $m_{jjj}$ markedly deviates from the global $K$-factor is crucial.

In the left panels of Fig.~\ref{fig-K-mass}, 
we exhibit the LO (in blue) and NLO (in orange) differential cross sections 
for the invariant mass distribution of three jets, alongside the $K$-factor distribution. 
The analysis focuses on the invariant mass window of $[0.8,2.0]$ TeV,
aiming at heavy new particles. 
Unlike the $p_T$ dependence, the invariant mass has a marginal impact on the $K$-factor, with deviations from the global value staying within approximately 10\%.

The second process we examine is $pp \to \gamma\gamma jj$. This process acts as a principal background for BSM Higgs decay scenarios into a pair of lighter new particles, such as a lighter Higgs boson or ALPs, which then decay into two photons\cite{ATLAS:2023ian,Wang:2023pqx}. Although the final state includes four photons, the background from four photons is negligible, with its total cross-section on the order of $10^{-5}\pb$. In contrast, the final state that consists of two photons and two jets (misidentified as photons) presents with a cross-section approximately 10 pb, thus becoming a significant background.

In the right panel of Fig.~\ref{fig-K-mass}, the LO (in blue) and NLO (in orange) differential cross-sections relative to $m_{\gamma\gamma jj}$, the invariant mass of two photons and two jets, are illustrated. Similar to the $pp \to jjj$ process, the deviation of the differential $K$-factor from the global $K$-factor across most $m_{\gamma\gamma jj}$ bins remains modest.

\section{Conclusions}
\label{sec:conclusions}

In this study, we have extensively analyzed the $K$-factors 
($K=\sigma_{\text{NLO}}/\sigma_{\text{LO}} \equiv 1+\delta K$) 
for a broad spectrum of SM processes at the 14 TeV LHC.
 Our analysis covers 111 processes,
 which are expected to serve as backgrounds for most BSM searches.  
Utilizing {\sc\small MadGraph5\_aMC@NLO} for our calculations,
we presented the LO cross sections alongside their corresponding $K$-factors.
We observed significant variation of $K$-factors across processes, 
from 1.001 ($pp \to jjj$) to 4.221 ($pp\to W^\pm \gamma\gamma\gamma$). 
This variance underscores the diverse impact of NLO corrections for different processes.
To provide a comprehensive overview, especially for processes with high $K$-factors, we also highlighted processes where $K$-factors exceed 1.5.

Key insights emerged from our analysis. 
Processes involving photons consistently showed exceptionally high $K$-factors,
mainly because of gluon-initiated processes at NLO that take advantage of the LHC's high gluon PDFs.
For instance, in the $pp \to \gamma W^\pm$ process, gluon-initiated real emissions account for about 40\% of the $K$-factor.
Moreover, processes featuring multiple particles, especially vector bosons, 
yielded high $K$-factors, a result of the interaction complexity and multiple real emission processes.  
An inverse correlation was also noted between the inclusion of jets and $\delta K$, indicating that adding jets typically reduces $\delta K$.

We also analyzed differential $K$-factors concerning transverse momentum 
and invariant mass, emphasizing their critical importance for BSM searches at the LHC.
The evaluation of differential $K$-factors for $p_T^j$ across various processes revealed significant increases with rising $p_T$, whereas the differential $K$-factors for invariant mass in selected processes of $pp\to jjj$ and $pp\to \gamma\gamma jj$ showed minimal deviation from global $K$-factors.

In conclusion, our extensive analysis underscores the necessity of accurately 
assessing the impact of high $K$-factors, particularly in the high $p_T$ region, 
on the BSM search. 
The findings from this comprehensive study are poised to guide future experimental strategies in the ongoing quest for new physics.

\section*{Acknowledgments}
This paper was supported by Konkuk University in 2023.

\end{document}